\documentclass[%
12pt,T
oneocolumn,
nofootinbib,
amsmath,amssymb,
aps,
floatfix,
unsortedaddress
]{revtex4-2}

\usepackage{graphicx,color}
\usepackage{subfig}
\usepackage{dcolumn}
\usepackage{bm}
\usepackage{graphicx,color}
\usepackage{float}
\usepackage{romannum}
\usepackage{amsmath}
\usepackage{mathtools}
\usepackage{upgreek}
\usepackage{cancel}
\usepackage{multirow}
\usepackage[pdfpagelabels, pdfencoding=auto, psdextra]{hyperref}
\hypersetup{%
	pdfsubject=Paper,
	pdfkeywords={nuclear physics} {Pionless EFT} {Two-Nucleon} {Lattice} {L\"usher formula},
	unicode = true,
	breaklinks = true,
	colorlinks = true,
	linkcolor = blue,
	citecolor = blue,
	menucolor = blue,
	citecolor = blue,
	urlcolor = blue
}

\graphicspath{{./}}
\begin{document}
	\pagenumbering{arabic}
	
	\title{  Exploring the $ \phi{\text -}\alpha $ interaction via femtoscopic study}
	\author{Faisal Etminan}
	\email{fetminan@birjand.ac.ir}
	\affiliation{
		Department of Physics, Faculty of Sciences, University of Birjand, Birjand 97175-615, Iran
	}%
	\affiliation{ Interdisciplinary Theoretical and Mathematical Sciences Program (iTHEMS), RIKEN, Wako 351-0198, Japan}

	\date{\today}%

	\begin{abstract}
Very recently the Wood-Saxon (WS) type interaction in the single-folding potential approach are constructed to simulate the $ \phi{\text -}\alpha $ potentials. 
One set of the $ \phi{\text -}\alpha $ potentials are based on the first principle HAL QCD  $\phi{\text -}N $ interactions in $^{4}S_{3/2}$ channel,
and in another set, the $\phi$-meson nucleus potentials were calculated by employing the quark-meson coupling (QMC) model. 
By utilizing  these two set of $ \phi{\text -}\alpha $ potentials, the  two-particle momentum correlation of $ \phi{\text -}\alpha $ in high-energy heavy ion collisions  is explored. 
The numerical results show that the correlation functions at small source size (high density nuclear medium) depends on the employed potential model. 
Also, the correlation functions are obtained within the Lednicky-Lyuboshits (LL) formalism. 
For small source size, it is found that, the LL  formula returns significantly different values due to the large interaction range of the $\phi{\text -}\alpha $ potential.
	\end{abstract}
	
	
	\maketitle
	\section{Introduction} \label{sec:intro}
	Although considerable experimental and theoretical~\cite{PhysRevC.95.055202,HIRENZAKI2010406,10.1143/PTP.124.147,Cobos2017prc,etminan2024prc,Filikhin2024prd,GUBLER2024100007} efforts have been done in the last few decades, the possible existence of $ \phi $-mesic bound states with
	nucleons (N) is still a challenging field of research.
	The essential basis for the studying of such a system is an exact information about the $\phi{\text -}N $ strong interaction.
	
	Up to now, no evidence of a $\phi{\text -}N $ bound state has been found, experimentally, since the standard invariant mass measurements of its decay products are disputing and the measurement  of 
	the $\phi{\text -}N $ scattering parameters are restricted to spin averaged quantities. A list of measurements that have been made for $ \phi $ absorption off different  nuclear targets is given in~\cite{CHIZZALI2024138358}, and the authors concluded that  the landscape of experimental results for the $\phi{\text -}N $ interaction is not convincingly in agreement and new information are required.  
	One method for understanding the hadron-hadron interaction (that is hard to investigate in scattering experiment) is measuring the 
	momentum correlation functions in high-energy collisions~\cite{cho2017exotic}.
	It can provides information on both the effective emission source and the interaction potential.
	Recent measurement of the correlation function of proton-$\phi$  in heavy-ion collisions by  ALICE collaboration  ~\cite{PhysRevLett.127.172301}, 
	together by indicating a $ p{-}\phi $ bound state using two-particle
	correlation functions~\cite{CHIZZALI2024138358,GUBLER2024100007,abreu2024studyphincorrelation} desire the proton${-}\phi$ bound state hypothesis.

	To go one step further, the femtoscopic analysis of hadron-nuclei  correlation functions
	could play a crucial role in understanding the structure and
	dynamics of the atomic nuclei~\cite{Morita2016,PhysRevC.108.064002,PhysRevX.14.031051}.
	Accordingly, the $ hd $ correlation functions are investigated for some hadron-deuteron systems like
	$ pd $   \cite{bazak2020production,PhysRevC.108.064002,mrowczynski2019hadron}, $ {K}^{-}d $ \cite{alicecollaboration2023exploring,mrowczynski2019hadron},
	$ \Lambda d $ \cite{Haidenbauerprc}, $ \Xi d $ \cite{Ogata2021},
	and $ \Omega NN $ ~\cite{zhang2021production,ETMINAN2023122639}.
	Furthermore, the momentum correlation between $ \Lambda \alpha $~\cite{jinno2024femtoscopic}, $ \Xi \alpha $ ~\cite{kamiya2024} and $ \Omega \alpha $ ~\cite{etminan2024omegaAlpha} have recently been studied theoretically.
	From experimental point of view, 
	the momentum correlation between (anti)protons and (anti)deuterons measured in pp collisions at $\sqrt{s} = 13$ TeV with ALICE is studied for the first time~\cite{Singh:2022r3} and  the  first measurement of $ d-\Lambda $ correlation with $ \sqrt{S_{NN}} =3 $ GeV Au+Au collisions is done by STAR Collaborations~\cite{hu2023}. 
	In addition, there is a research plan at the Thomas Jefferson National Accelerator Facility (JLab), to investigate
	the binding of $ \phi $ (and $\eta $) to $^{4}He (\alpha) $~\cite{Cobos2017prc}.
	

	A recent   HAL QCD simulations of the $ \phi{\text -}N $ potential
	in $^{4}S_{3/2}$ channel (by the representation $^{2s+1}L_{J}$ that $s, L$ and
	$J$ are the total spin, orbital angular momentum and total angular momentum,
	respectively) are performed~\cite{yan2022prd}. 
	The simulation is done  on the $\left(2+1\right)$-flavor
	with nearly physical quark masses  $\left(m_{\pi},m_{K}\right)\simeq\left(146,525\right)$ MeV on a large lattice volume of 
	$\simeq\left(8.1\:{fm}\right)^{3}$ at the imaginary-time slices $ t/a=14 $ where $ a=0.0846 $ fm is the lattice spacing. 
	In Ref.~\cite{filikhin2024phihe} the $ \phi{\text -}\alpha $ potential was obtained through a single folding of the HAL QCD $ \phi{\text -}N $ interaction in the $ ^{4}S_{3/2} $ channel with the nuclear matter distribution of $ ^{4}He $. Moreover, a Wood-Saxon (WS) type interaction  was built to simulate the $ \phi{\text -}\alpha $ potential, given in ~\cite{Cobos2017prc}, based on an effective Lagrangian approach where the nuclear density distributions, as well as the in-medium $ K $ and $\bar{K} $ meson masses, were consistently calculated in the frame work of
	quark-meson coupling (QMC) model~\cite{SAITO20071}. In Ref.~\cite{filikhin2024phihe}, the authors concluded that there were qualitatively a good agreement between binding energies for both types of the $ \phi{\text -}\alpha $ interactions.

	Therefore, motivated by the above discussions, in this work, I want to explore the $ \phi{\text -}\alpha $ correlation function in the relativistic heavy ion collisions to probe the nature of $ \Omega N $ interactions as an independent source of information. 
	The purpose of this work is to give an illustration for what can be expected from measuring $ \phi{\text -}\alpha $ correlations.
	Since this is an  exploratory study, the techniques used are simple.

	The paper is organized in the following way:
	In Sec.~\ref{sec:featur-of-potential},  models of the $ \phi{\text -}\alpha $ potential, HAL QCD and QMC  are introduced, and discussed their behaviour and properties. In Sec.~\ref{sec:Two-particle-CF} the formalism for two-particle momentum correlation
	functions is briefly reviewed. 
	The summary and  conclusions  are given in Sec.~\ref{sec:Summary-and-conclusions}.  
	\section{Features of $ \phi{\text -}\alpha $ potentials   } \label{sec:featur-of-potential}	
	The single folding approach approximates the effective $ \phi+\alpha $ nuclear potential from two body $\phi{\text -}N$ potential,
	$V_{\phi{\text -}N}\left(\left|\textbf{r}-\textbf{r}^{\prime}\right|\right)$ as
	~\cite{Satchler1979,Miyamoto2018,Etminan:2019gds}
	\begin{equation}
		U_{\phi{\text -}\alpha}\left(r\right)=\int\rho\left(r^{\prime}\right)V_{\phi{\text -}N}\left(\left|\textbf{r}-\textbf{r}^{\prime}\right|\right)d\textbf{r}^{\prime},\label{eq:V_alfaphi}
	\end{equation}		
	where $\rho\left(r^{\prime}\right)$ is the nucleon density distribution in
	$ ^{4}He $-particle at a distance $\textbf{r}^{\prime}$ from its
	center-of-mass.
	The detailed calculations of how to obtain the $ U_{\phi{\text -}\alpha}\left(r\right) $ is given by Filikhin et al. in Ref.~\cite{filikhin2024phihe}. Here, I just mention the main part. 
	As follows from Ref.~\cite{PhysRevC.109.L012201}, Filikhin et al. chose the simple Gaussian matter density distribution model as
	\begin{equation}
		\rho\left(r\right)=\left(\frac{C^{2}}{\pi}\right)^{3/2}\exp\left(-\left(Cr\right)^{2}\right),
	\end{equation}
	that reproduces the experimental root-of-mean-square (rms) radius,
	$ \sqrt{\left\langle r^{2}\right\rangle }=\sqrt{\frac{3}{2C^{2}}} =1.70 \pm 0.14 $~\cite{PhysRevC.109.L012201}. 
	Therefore, the numerical calculations are done for three values of rms radius, i.e, $ 1.56 $ fm, $ 1.70 $ fm, and $ 1.84 $ fm, 
	in order to study the influence of the rms on  $ \phi{\text -}\alpha $ potential in~\cite{filikhin2024phihe}.
	
\begin{figure*}[hbt!]
		\centering
		\includegraphics[scale=1.0]{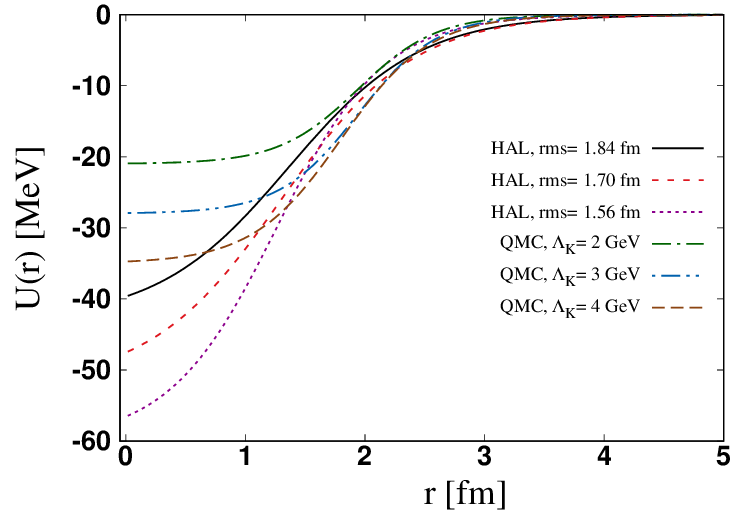}
		\caption{ The single-folding potentials $U_{\phi{\text -}\alpha}\left(r\right)$,  as functions of distance between $ \phi $ and $\alpha$ by parameters given in Ref.\cite{filikhin2024phihe}. 
			The first three based on  HAL QCD  ($ t/a=14 $) $ \phi{\text -}N $ potentials for the three values of the rms radius: $ 1.84 $ fm (solid black line), $ 1.70 $ fm (short dashed red line), and $ 1.56 $ fm (dotted magenta line). 
			And the second three based on the QMC model of $ \phi{\text -}N $ potentials for three values of the cutoff parameter $ \Lambda_{K} = 2 $ GeV (dash-dotted green line), $ 3 $ GeV (dash-dot-dotted blue line) and $ 4 $ GeV (dashed brown line), where the rms matter radius of $ ^{4}He $ is chosen to be $ 1.56 $ fm.		
			The corresponding parameters are summarized in table~\ref{tab:phialpha-para}. 		
			\label{fig:v_phiAlpha}}
	\end{figure*}

	In Ref.~\cite{filikhin2024phihe},
	for phenomenological application and calculation of observables, such as scattering phase shifts and binding energies, the $U_{\phi{\text -}\alpha}\left(r\right)$ is fitted to a Wood-Saxon form using the function that given by Eq.~\eqref{eq:ws-fit} (motivated by common Dover-Gal model of potential~\cite{dover1983}),			
	\begin{equation}
		U^{fit}_{\phi{\text -}\alpha}\left(r\right)=-V_{0}\left[1+\exp\left(\frac{r-R}{c}\right)\right]^{-1} , \label{eq:ws-fit}
	\end{equation}		
	where the depth parameter $ V_{0}$ indicates the strength of interaction, $ R = 1.1 A^{1/3}$ where $ A =4 $ is the mass number of the $ \alpha $ nucleus and $ c $ shows the surface diffuseness.
	The obtained value of the parameters of Eq.~\eqref{eq:ws-fit}  correspond to these three rms radius are given in the table~\ref{tab:phialpha-para} and illustrated in Fig.~\ref{fig:v_phiAlpha}. Note that concrete parameterizations are taken straight from Ref.~\cite{filikhin2024phihe}.
	
	\begin{table}[hbt!]
		\caption{
			The parameters of $\phi{\text -}\alpha$ potential in Eq.~\eqref{eq:ws-fit} based on HAL QCD  (at lattice Euclidean time  $ t/a=14 $) and quark-meson coupling model of $ \phi{\text -}N $ potentials.  The parameters are taken straight from Ref.~\cite{filikhin2024phihe}.
			Also, scattering length $ a_{0} $, effective range $r_{0}$ and binding energy $B_{\phi{\text -}\alpha}$ are presented.
			The results have been obtained by employing the experimental masses of $ m_{\alpha} =3727.38 $ MeV/c and  $m_{\phi}=1019.5$  MeV/c . 
			\label{tab:phialpha-para}}	
		\begin{tabular}{ccccccccc}
			\hline
			\hline 
			Model &            &$ rms $ (fm) &$ V_{0} $ (MeV) & $ R $ (fm)& $ c $ (fm)& $a_{0}$ (fm)&$r_{0}$ (fm)& $B_{\phi{\text -}\alpha}$ (MeV)\\
			\hline
			\multirow{ 3}{*}{$U_{\phi{\text -}\alpha}^{HAL}$ }
			&                  & $ 1.84 $& $43$& $ 1.36$&$ 0.55$&$4.0 $ & $1.7$ & $2.97 $ \\
			&                  & $ 1.70 $& $52$& $ 1.30$&$ 0.55$&$3.4 $ & $1.5$ & $4.78 $ \\
			&                  & $ 1.56 $& $60$& $ 1.26$&$ 0.45$&$3.0 $ & $1.3$ & $5.98 $ \\
			\hline
			\multirow{ 3}{*}{$ U_{\phi{\text -}\alpha}^{QMC} $}
			& $ \Lambda_{K}=2 $ (GeV)& $ 1.56 $& $21$& $ 1.94$&$ 0.33$&$6.7$ & $2.0$ & $0.80 $ \\
			& $ \Lambda_{K}=3 $ (GeV)& $ 1.56 $& $28$& $ 1.94$&$ 0.33$&$3.9$ & $1.7$ & $3.19 $ \\
			& $ \Lambda_{K}=4 $ (GeV)& $ 1.56 $& $35$& $ 1.80$&$ 0.37$&$3.4$ & $1.6$ & $4.71 $ \\	
			\hline
			\hline 	
		\end{tabular}
	\end{table}
	
	In Ref.~\cite{Cobos2017prc}, the $ \phi $-meson–nuclear potentials were calculated employing a local
	density approximation, by evaluating the $ K \bar{K} $ loop contribution in the $ \phi $ self-energy. 
	The nuclear density distributions, as well as the in-medium $ K $ and $\bar{K} $ meson masses, were consistently calculated in the frame work of quark-meson coupling (QMC) model~\cite{SAITO20071}. 
	Furthermore, they investigated the sensitivity of results to the cutoff parameter in the form factor at the $ \phi{\text -} K\bar{K} $ vertex appearing in the $ \phi $-meson self-energy. 
	The resultant $ \phi{\text -} \alpha $ potential  is given for three values of the cutoff parameter, i.e., $ \Lambda_{K} = 2, 3$ and $4$ GeV.

	Accordingly, Filikhin et al. in Ref.~\cite{filikhin2024phihe}, constructed a Wood-Saxon (WS) type  $ \phi{\text -}\alpha $ potentials to simulate  $ \phi{\text -} \alpha $ potentials for  $ \Lambda_{K}  =2, 3 $ and $ 4 $ GeV.
	In Fig.~\ref{fig:v_phiAlpha}, for comparison, I just represent the $ \phi{\text -}\alpha $ potentials, denoted as
	$ U_{\phi{\text -}\alpha}^{QMC} $ by using the parameters given in~Ref.~\cite{filikhin2024phihe} directly, for three values of the
	cutoff parameter $ \Lambda_{K} = 2, 3$ and $ 4 $ GeV. 
	In table~\ref{tab:phialpha-para}, the parameters of $\phi{\text -}\alpha$ potential in Eq.~\eqref{eq:ws-fit} based on HAL QCD  (at lattice Euclidean time  $ t/a=14 $) and quark-meson coupling model of $ \phi{\text -}N $ potentials, are given.  
	
	From Fig.~\ref{fig:v_phiAlpha}, it is clear that the depth of the  potential is sensitive to the cutoff parameter, varying from  $-20 $ MeV to $-35 $ MeV.
	However, the bound state energy is obviously dependent on $ \Lambda_{K} $, increasing as $ \Lambda_{K} $ increases (see $B_{\phi{\text -}\alpha}$'s value in table~\ref{tab:phialpha-para}).
	The attractive potential for the $ \phi $ meson in nuclear medium originates from the in-medium enhanced $ K\bar{K} $ loop in the $ \phi $-meson self-energy~\cite{Cobos2017prc}.
	According to the Fig.~\ref{fig:v_phiAlpha} and table~\ref{tab:phialpha-para},  
	the obtained  $ \phi{\text -}\alpha $ potentials based on $ \phi{\text -}\alpha $ HAL potential, i.e., $ U_{\phi{\text -}\alpha}^{HAL} $, is more  attractive  than $ U_{\phi{\text -}\alpha}^{QMC} $  almost at all distances. The former is much deeper than latter and more rapidly goes to zero. But, in both cases, the interaction ranges are about $ 3 $ fm, where  potentials become almost zero.

\begin{figure*}[hbt!]
		\centering
		\includegraphics[scale=1.0]{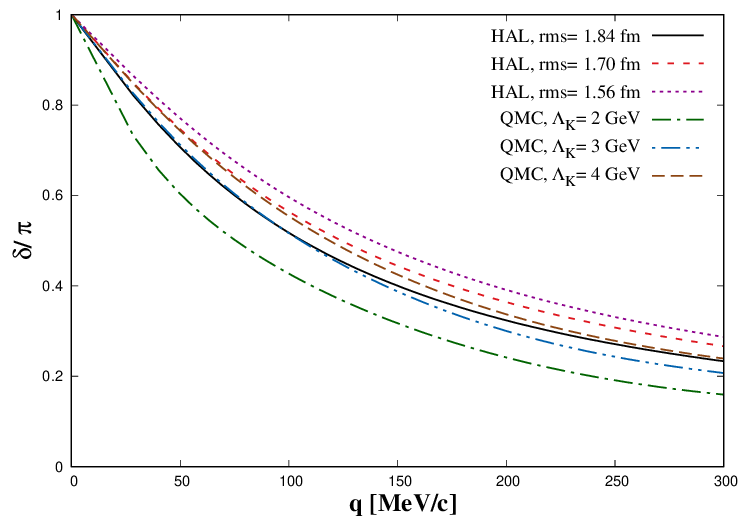}
		\caption{ 
			The normalized $\phi{\text -}\alpha$ scattering phase shifts $ \delta /\pi$  as functions of the relative momentum $ q $  with model of HAL QCD for the three values of the rms radius: $ 1.84 $ fm (solid black line), $ 1.70 $ fm (short dashed red line), and $ 1.56 $ fm (dotted magenta line)  and  QMC  for three values of the cutoff parameter $ \Lambda_{K} = 2 $ GeV (dash-dotted green line), $ 3 $ GeV (dash-dot-dotted blue line) and $ 4 $ GeV (dashed brown line).  
			\label{fig:phase_DG_HAL}}
	\end{figure*}

	By employing $ \phi{\text -}\alpha $ potentials as input, the Schr\"{o}dinger equation is solved to extract binding energy and  scattering observables from the asymptotic behavior of the wave function.
	Fig.~\ref{fig:phase_DG_HAL} depicts the normalized $\phi{\text -}\alpha$ phase shifts $ \delta /\pi$ calculated as functions of the 
	magnitude of the relative momentum $ q=\sqrt{2\mu E} $  ($ \mu $ is the reduced mass of $ \phi{\text -}\alpha $ system) with model of HAL QCD for the three values of rms  and  QMC potentials for the three values of $ \Lambda_{K} $. 
	The phase shift behaviour for the all case represents an attractive interaction even to form a bound state.  
	The scattering length $ a_{0} $ and the effective range $r_{0}$  are defined by using the effective range expansion (ERE) formula up to the next-leading-order (NLO),		
	\begin{equation}
		q \cot\delta=-\frac{1}{a_{0}}+\frac{1}{2}r_{0}q^{2}+\mathcal{O}\left(q^{4}\right).\label{eq:ERE}
	\end{equation}		
	The corresponding results for all cases are given in table~\ref{tab:phialpha-para}. 
	
	\section{Two-particle correlation function}\label{sec:Two-particle-CF}
	The correlation function formulae has been described in many work like~\cite{KOONIN197743,Pratt1986,PhysRevC.91.024916,OHNISHI2016294,cho2017exotic}. We bring here only the key formulae. 
	Two-particle momentum correlation function $C_{q}$ for a given $\phi{\text -}\alpha$ potential can be obtained by Koonin-Pratt (KP) formula~\cite{OHNISHI2016294}
	\begin{equation}
		C\left(q\right)=\int4\pi r^{2}drS\left(\boldsymbol{r}\right)\left|\Psi_{\phi{\text{-}}\alpha}^{\left(-\right)}\left(\boldsymbol{r},\boldsymbol{q}\right)\right|^{2}, \label{eq:kp}
	\end{equation}
	$ S\left(r\right) $  is the properly normalized single particle source function and in this work, it is supposed to be spherical and static Gaussian with source size $R$, 
	$ S\left(r\right)=\exp\left(-r^{2}/4R^{2}\right)/\left(2\sqrt{\pi}R\right)^{3} $.
	$\Psi_{\phi{\text{-}}\alpha}^{\left(-\right)} $ is the relative wave function with the outgoing boundary condition. 
	That can be calculated easily for a two-body systems by solving the  Schr\"{o}dinger equation for a given  $ \phi{\text -}\alpha $ potential.
	
	The calculations of two-particle correlations via the KP formula, Eq.~\eqref{eq:kp}, is done by employing the "\textit{Correlation Analysis tool using the Schr\"{o}dinger Equation}" (CATS)~\cite{cats}. 
	For given an interaction potential and an emission source CATS tool kit computes the correlation function~\cite{mihaylov2023novel}.

\begin{figure*}[hbt!]
		\centering
		\includegraphics[scale=0.64]{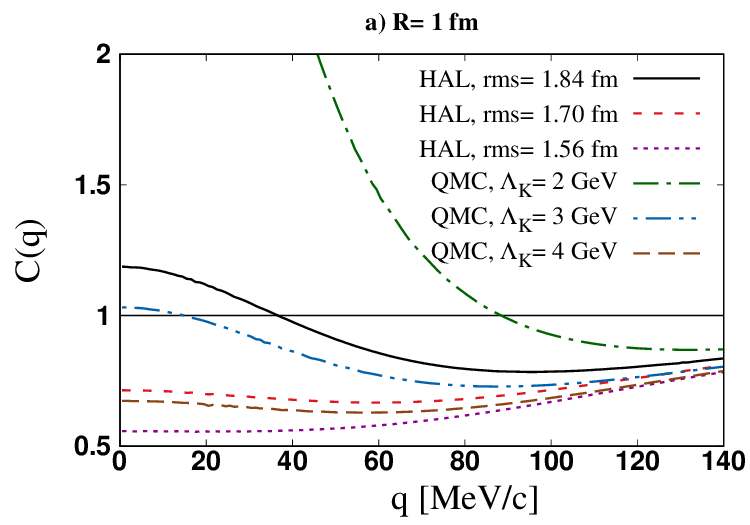}    
		
		\includegraphics[scale=0.64]{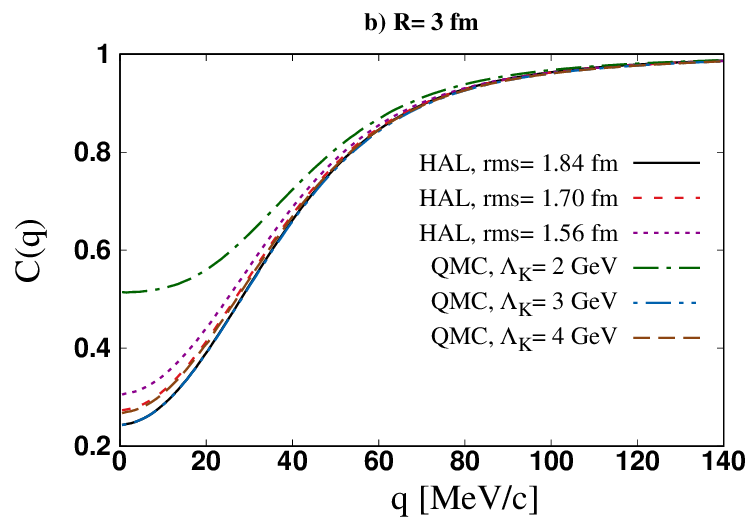} \includegraphics[scale=0.64]{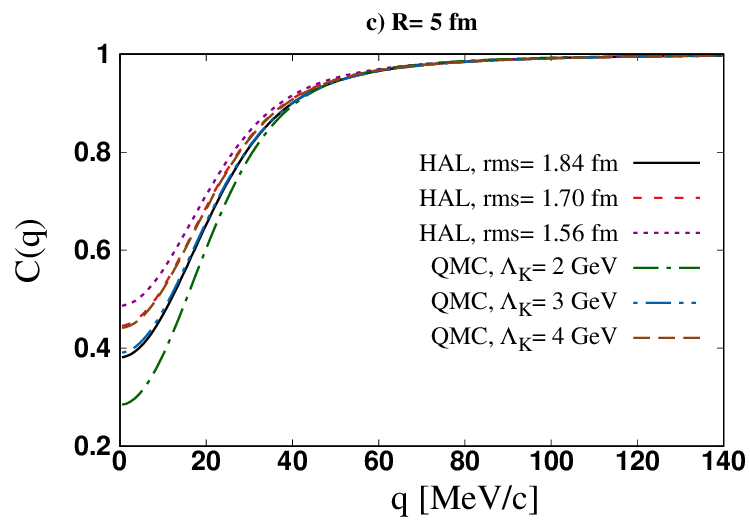}
		\caption{ The $ \phi{\text -}\alpha $ correlation functions for three different source sizes: a) $ R=1 $ fm, b) $ R=3 $ fm and c) $ R=5 $ fm, 
			with model of HAL QCD for the three values of the rms radius: $ 1.84 $ fm (solid black line), $ 1.70 $ fm (short dashed red line), and $ 1.56 $ fm (dotted magenta line)  and  QMC  for three values of the cutoff parameter $ \Lambda_{K} = 2 $ GeV (dash-dotted green line), $ 3 $ GeV (dash-dot-dotted blue line) and $ 4 $ GeV (dashed brown line).
			\label{fig:phiAlpha_cq_kp_R}}
	\end{figure*}

	The $ \phi{\text -}\alpha $ correlation functions by the KP formula, Eq.~\eqref{eq:kp} for three different source sizes, 	$ R = 1, 3 $ fm and $ 5 $ fm  are calculated with various $ \phi{\text -}\alpha $ potentials and depicted in Fig.~\ref{fig:phiAlpha_cq_kp_R} a) , b) and c), respectively.	
	The values of source sizes in accord with the typical sizes applied in
	the exploration of two-hadron correlation functions	~\cite{jinno2024femtoscopic,kamiya2024}.
	The particular dip shape can be observed at small source size $ R = 1 $ fm, that is conventional in the bound state of system near the threshold~\cite{jinno2024femtoscopic} (except  in the  case of QMC potential model by cutoff parameter $ \Lambda_{K} = 2 $ GeV). 	
	Moreover, Fig.~\ref{fig:phiAlpha_cq_kp_R} (a) shows the results for all cases of two potentials are different at low momentum $ q\lesssim100 $ MeV/c.		
	According to Fig.~\ref{fig:v_phiAlpha} the HAL potential model, is more attractive than QMC potential model, thus it gives enhancement of $ C_{\phi\alpha}\left(q\right) $. 
	
	At a larger source size $ R = 3 $, the QMC potential model by cutoff parameter $ \Lambda_{K} =2 $ GeV
	are distinguishable by the measurement of the $ \phi{\text -}\alpha $ correlation functions.
	Nevertheless, with the increase of the source size, the difference between 
	the $ C_{\phi\alpha}\left(q\right) $s decreases until they are almost same for $ R = 5 $ fm.
	As a results, in the future experiments, the measurement of $ \phi{\text -}\alpha $ correlation function from
	a small source at relatively low momentum, significantly can be imposed by $ \phi{\text -}\alpha $ interaction at high densities.
	
	In the two-body systems if the interaction range is much smaller than source size, it is possible to further investigate 
	the interaction dependence of the correlation function	by means of
	Lednicky-Lyuboshits (LL) formula~\cite{Lednicky:1981su,OHNISHI2016294},
	\begin{equation}
		C_{LL}\left(q\right)=1+\frac{\left|f\left(q\right)\right|^{2}}{2R^{2}}F_{0}\left(\frac{r_{0}}{R}\right)+\frac{2\textrm{Re}\:f\left(q\right)}{\sqrt{\pi}R}F_{1}\left(2qR\right)-\frac{\textrm{Im}\:f\left(q\right)}{R}F_{2}\left(2qR\right), \label{eq:ll}
	\end{equation}
	$f\left(q\right)\approx1/\left(-1/a_{0}+r_{0}q^{2}/2-iq\right)$ is  scattering amplitude and it is evaluated from
	the  ERE formula Eq.~\eqref{eq:ERE}.
	In addition, $F_{1}\left(x\right)=\int_{0}^{x}dt\:e^{t^{2}-x^{2}}/x,\:F_{2}\left(x\right)=\left(1-e^{-x^{2}}\right)/x,$
	and $F_{0}\left(x\right)=1-x/\left(2\sqrt{\pi}\right)$~\cite{Lednicky:1981su,OHNISHI2016294} are mathematical functions.      
	Since the LL relation is extracted from the KP relation 
	by approximating the full wave function by the asymptotic wave function and the effective range correction,
	by comparing the correlation function via the KP and the LL formulae
	it is possible to decode the detailed shape of potential.

\begin{figure*}[hbt!]
		\centering
		\includegraphics[scale=0.64]{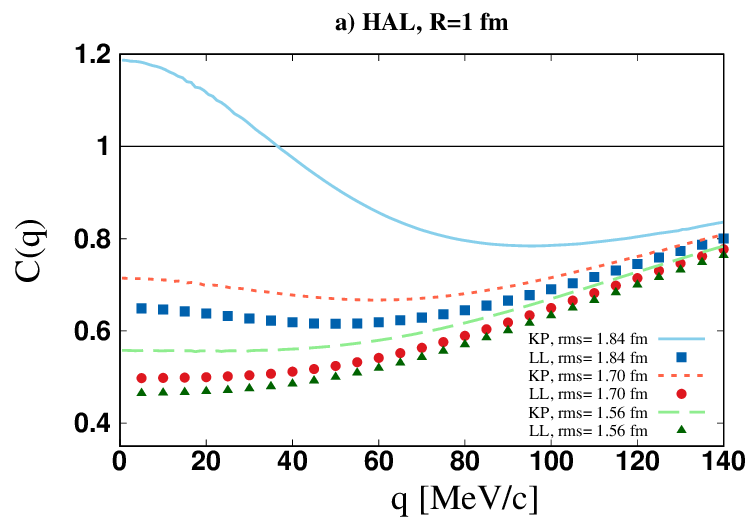} \includegraphics[scale=0.64]{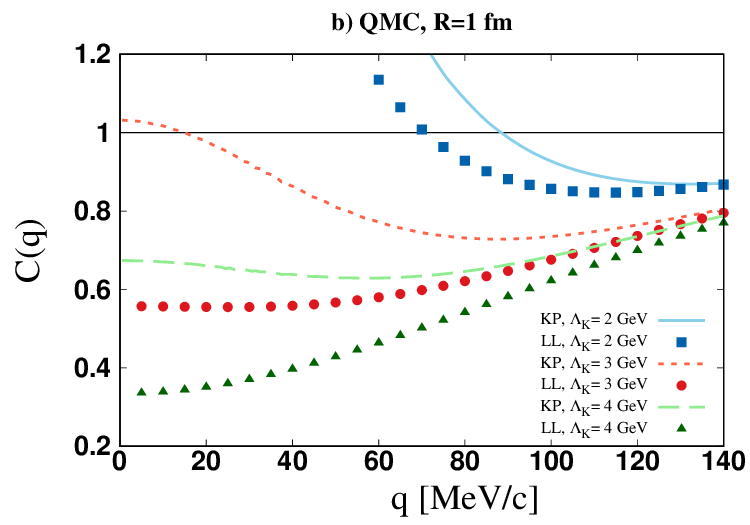}
		\includegraphics[scale=0.64]{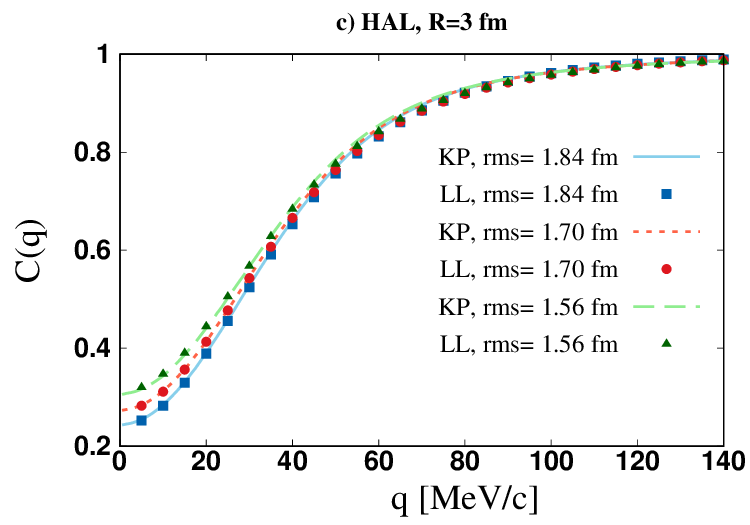} \includegraphics[scale=0.64]{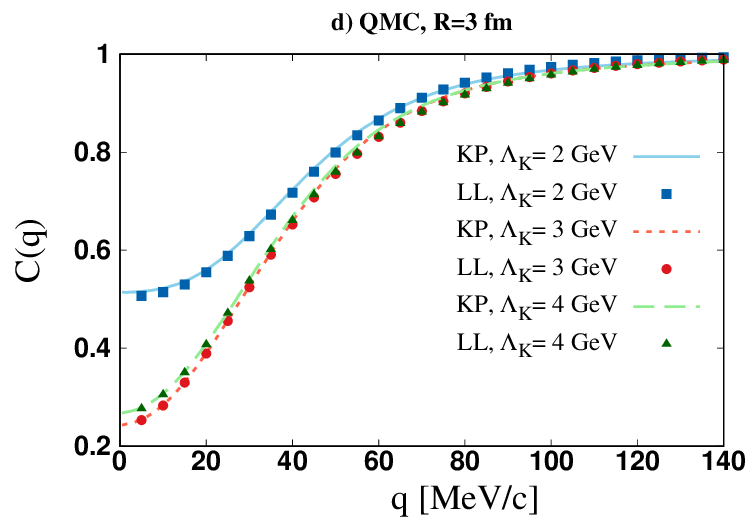}
		\includegraphics[scale=0.64]{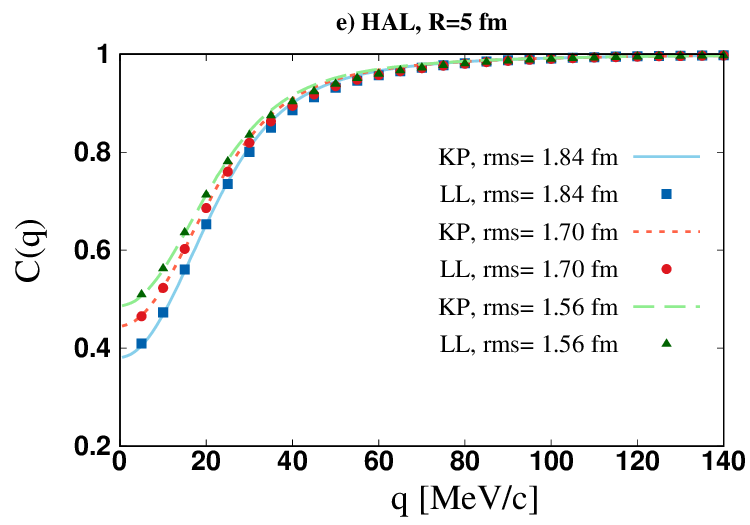} \includegraphics[scale=0.64]{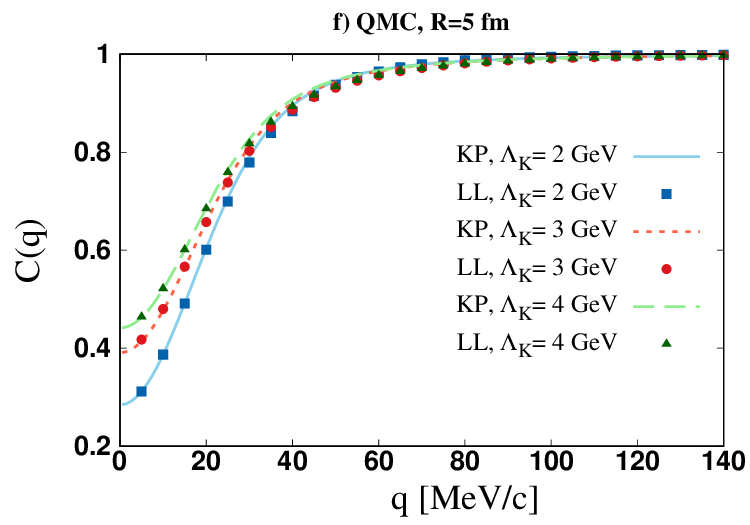}
		\caption{ 
			Comparison of $ \phi{\text -}\alpha $ correlation functions by HAL (Left panels a, c, e) and QMC model of potentials (Right panels b, d, f) with the
			The KP formula  and the LL formula estimation for three source sizes $ R=1, 3 $ and $ 5 $ fm.
			The results are shown for  HAL QCD  with three values of the rms radius: $ 1.84 $ fm (blue solid line and filled square), 
			$ 1.70 $ fm (red dotted line and filled circle), and $ 1.56 $ fm (green dashed line and filled triangle)  
			and  
			for QMC  model  with three values of the cutoff parameter $ \Lambda_{K} = 2 $ GeV (blue solid line and filled square), $ 3 $ GeV (red dotted line and filled circle) and $ 4 $ GeV (green dashed line and filled triangle).  
			\label{fig:phiAlpha_cq_kp_ll_R}}
	\end{figure*}

	$ \phi{\text -}\alpha $  correlation functions is estimated via  the LL formula (Eq.~\ref{eq:ll}) using 	
	the scattering length and the effective range of two models of potentials given in table~\ref{tab:phialpha-para}. 
	And, the results are compared with the ones from the KP formula in Fig.~\ref{fig:phiAlpha_cq_kp_ll_R} for three different source sizes $ R=1, 3 $ and $ 5 $ fm.  
	Figure ~\ref{fig:phiAlpha_cq_kp_ll_R} a) and b) show the results of calculations by both potentials HAL (left panel) and QMC (right panel) models for $ R=1 $ fm source size. They reveal that, the LL approach generates significant different results of the KP formula at low momentum region.
	In reactions involving light nuclei, the interaction range usually is $ \gtrsim 3  $ fm~\cite{mrowczynski2019hadron,bazak2020production,StanislawPRC2021,jinno2024femtoscopic}.
	Therefore, the LL formula is an incapable approximation where the source size
	is smaller than the interaction range~\cite{OHNISHI2016294}.
	Moreover, it can be seen that there is a different behavior about the QMC potential model by cutoff parameter $ \Lambda_{K}=2 $ GeV 
	which according to the figure~\ref{fig:v_phiAlpha} and data in table~\ref{tab:phialpha-para} is the shallowest potential.
	While the correlation function for another models of potentials  have almost same behaviour.
	So, it is possible to reveal the difference between potentials by precision measurement of correlation function at almost small source size $R=1 \sim 3$ fm.
	For  larger source size $ R =5 $ fm, the figure ~\ref{fig:phiAlpha_cq_kp_ll_R} e) and f)  show that the correlation functions for the LL formula are in acceptable agreement by the KP formula ones.
	\section{Summary and conclusions\label{sec:Summary-and-conclusions}}	
	Very recently, six Wood-Saxon type potential through a single-folding potential approach are constructed  to
	simulate the $ \phi{\text -}\alpha $ potentials. 
	Three of them are made by using the  first principle HAL QCD  $\phi{\text -}N $ interactions in $^{4}S_{3/2}$ channel 
	and a Gaussian form of the $\alpha$-particle matter distribution 
	for three values of the rms radius,  $ 1.84, 170 $  and $ 1.56 $ fm, where these values are motivated by the experimental measurement~\cite{PhysRevC.109.L012201}.
	And conversely, another three of them,  are obtained 
	based on the attractive potential for the $\phi$-meson in the nuclear medium,  
	that calculated by employing the QMC model for three values of the cutoff parameter, $ \Lambda_{K} = 2, 3 $  and $ 4 $ GeV. 		
	Roughly, the $ \phi{\text -}\alpha $ potentials based on HAL  is more attractive and deeper than one based on QMC model.  In both cases, the interaction ranges are about $ 3 $ fm. 
	
	In this work, 	I employed femtoscopy technique to predict $ \phi{\text -}\alpha $ momentum correlation functions in
	the high-energy collisions looking for an additional and alternative source of knowledge relevance to the
	$ \phi N $ interaction by using these six different type of $ \phi{\text -}\alpha $ potentials.  
	Employing two $ \phi{\text -}\alpha $ potentials, correlation functions are calculated  using the KP formula for three different source sizes, $ R=1,3 $ fm and $ 5 $ fm. 
	The difference of  potentials is appear in the correlation functions by small source size around $ 1-3 $ fm, while for source size $ R\gtrsim5 $ fm the correlation functions tended to become same for both $ \phi{\text -}\alpha $ potentials. 
	In conclusion, since the correlation functions are sensitive to $ \phi{\text -}\alpha $ potentials behavior, we could get important information about  effects of $ \phi $ particle in dense nuclear medium.
	
	Furthermore, 
	the  scattering length and effective range were calculated by solving the
	Schr\"{o}dinger equation for each of the potentials separately. 
	Next, correlation functions is examined within the Lednicky-Lyuboshits approach  and compared by the results of the KP formula. 
	It was seen as expected, the LL formula by small source ($ 1 $ fm) seriously differ from the KP formula in the low-momentum region.
	
	Last but not least, in this exploratory study, the calculation are done for source sizes $ R = 1, 3 $, and $ 5 $ fm. 
	where the choice is motivated by values suggested by analyses of measurements of the two-hadron correlation function in $ pp $ collisions and heavy ion collisions~\cite{jinno2024femtoscopic,kamiya2024}.
	I note that the Koonin-Pratt formula Eq.~\eqref{eq:kp} is valid as long as the two correlated particles can be treated as point-like objects, i. e. the correlated particles must be well separated. 
	The  source size of the composite particle $ \alpha $ should  be bigger than those with single hadron emissions,  
	because there is a possibility of $ \alpha$ particle formation at the same time~\cite{mrowczynski2019hadron,bazak2020production,StanislawPRC2021}.
	Then, we rather deal with a five-body problem of two protons, two neutrons and $ \phi $ and a formation of alpha particle and creation of $ \phi{\text -}\alpha $ correlation occur simultaneously. 	
	Practically, this point should be considered in future works and is beyond the scope of this paper.	
	I hope that  theoretical results for $ \phi{\text -}\alpha $ momentum correlation functions together by 
	future experiments at  FAIR~\cite{cbm}, NICA, and J-PARC HI~\cite{j-park} sheds light on $ \phi{\text -}N $ interactions.
	
	\section*{Acknowledgement}
	I am grateful for the authors and maintainers of "\textit{Correlation Analysis tool using the Schr\"{o}dinger Equation}" (CATS)~\cite{cats}, a modified version of which is used for calculations in this exploratory study.	
	Discussions during the long-term workshop, HHIQCD2024 at Yukawa Institute for Theoretical Physics (YITP-T-24-02), were useful as I finished this work.		

	
	\bibliography{Refs.bib}
	
\end{document}